\documentclass[twoside]{LCWS11}
\usepackage[latin1]{inputenc}
\usepackage[dvips]{graphicx,epsfig,color}
\usepackage{wrapfig,rotating}
\usepackage{amssymb,amsmath,array}
\usepackage{cite}
\begin{document}
\title{
Photon Linear Collider Gamma-Gamma Summary} 
\author{Jeff Gronberg
\thanks{This work performed under the auspices of the U.S. Department of Energy by the Lawrence Livermore National Laboratory under Contract DE-AC52-07NA27344}
\vspace{.3cm}\\
Lawrence Livermore National Laboratory\\
L-050, 7000 East Ave. Livermore, CA, 94550 - USA
}

\maketitle

\begin{abstract}
High energy photon - photon collisions can be achieved by adding high average
power short-pulse lasers to the Linear Collider, enabling an expanded
physics program for the facility.
The technology required to realize a photon linear collider continues to
mature.
Compton back-scattering technology is being developed around the world for
low energy light source applications and high average power lasers are
being developed for Inertial Confinement Fusion.
\end{abstract}

\section{Physics Case}

The physics case for a high energy photon linear collider has been previously 
reviewed in~\cite{Bechtel2006}.  While there is a variety of interesting
processes, many of the most interesting 
processes involve the direct production of a Higgs boson.
The Higgs boson can be produced in photon-photon collisions through a triangle 
diagram which is sensitive to the existence of charged particles 
of arbitrarily high mass.  This single particle production allows for 
a greater energy reach for the production of supersymmetric Higgs bosons 
(MSSM H and A) in regions of low sensitivity at the LHC.  Control of the 
polarization of the Compton photons allows intial states of definte CP to
be created providing a probe of the CP nature of the observed Higgs boson.

The photon linear collider also provides an opportunity to search for 
new physics
by searching for anomalous couplings in double and single W boson production.
As well there is the potential for production of supersymmetric 
particles in electron-gamma 
collisions.  This would provide a greater energy reach to search for 
these particles than in the linear collider.

\section{Technology Requirements}

The fundamental physics of Compton scattering and the design of a 
photon collider has been laid out~\cite{Telnov1981,ECFA01}. 
Compton scattering can transfer 80\% of the incident electron energy to 
the backscattered photons when a 1~micron wavelength laser pulse is scattered
from a 250~GeV electron beam.  A laser pulse of 5~Joules, compressed to 1~ps
width and focused to a diffraction limited spot can convert most of the
incoming electrons in a bunch to high energy photons.  An enormous amount
of average laser power would be required to provide 15,000 laser
pulses per second to match the electron beam structure.  Since most of the
laser energy goes unused in the Compton process the required energy can
be greatly reduced if the laser pulses can be recirculated.

A design of a recirculating cavity~\cite{Will2001} was created in 
2001 which takes advantage
of the long inter-bunch spacing in the superconducting machine to recirculate
the laser pulses around the outside of the detector.  Calculations showed
that the required laser power could be reduced by a factor of 300 in this
design.  Recent studies have shown that a laser with sufficient phase 
stability to drive such a cavity is probably ahievable with current technology
and would cost in the vicinity of 20 million dollars.

Warm RF technology has a different electron bunch structure which precludes
use of the large recirculating cavity.  The small inter-bunch spacing does
not allow sufficient time for laser pulses to travel around the outside of the
detector.
A concept for a linear collider based on 
warm RF systems could be either a single pass system or a much smaller 
recirculating cavity within the detector with 
multiple circulating bunches.  A single pass system would still require
a prodigous amount of average laser power.  Capital costs for such a system
would be driven by the cost of laser diodes to transfer wall plug power 
to the laser amplifier crystals.  The available power saving for a 
recirculating system would depend on the achievable size of the cavity.
That would determine how many times a pulse could be reused in a single 
electron bunch train.

Implementation of the photon collider option has several requirements for
both the detector and the electron accelerator.  Apertures must be opened
in the forward part of the detector to allow the laser pulses to reach
the Interaction Point and be focused a few millimeters before the electron
beams collide.  
The electron beam will be left with an enormous energy spread after
the Compton backscatter and a large crossing angle will be required in 
order to allow sufficient aperture for the spent beam to be extracted.  
Finally, the photon collider option will require its own beam dump design
in order to handle the photon beam which will have about 50\% of the
final beam energy.

\section{Ongoing Developement of Compton Technology}

Compton backscattering for the creation of MeV gamma-ray light sources is
a world-wide activity.  The basic techniques of bringing an electron beam
and a laser pulse into collision is independent of the electron beam energy
and these facilities are providing vital experience in the development of
these techniques for the linear collider.  These facilities are also
developing the technology for recirculating laser pulses which will be critical
to achieve a cost effective solution for the photon linear collider.

Current MeV gamma-ray sources include the ThomX~\cite{Variola2011}
 machine at LAL, the LUCX~\cite{Fukuda2010} 
machine at KEK and the T-REX~\cite{Hartemann2010} machine at LLNL.  
The MightyLaser collaboration
is developing a four mirror recirculating cavity for the demonstration of
Compton backscattering at ATF~\cite{Delerue2011}.

\section{Future Prospects}

The technology of Compton backscattering and recirculating cavities is
being developed for the applications of MeV light sources, polarized 
positron sources and beam diagnostics.  High average power, short-pulse 
lasers are being developed for the application of Inertial Confinement
Fusion.  All of these efforts push forward the technical feasibility of
a photon linear collider.  While the photon linear collider has always
been envisioned as a second stage to the basic linear collider program
there may be advantages to considering it as a first stage.  The photon
collider requires an electron linear collider to drive it but it does not
require positrons and it does not require flat electron beams at the
Interaction Point in order to reduce the beamsstrahlung.  This opens up the
possibility of creating a first stage linear collider without a positron
source.  The creation of a low-emittance RF electron gun might also create
the possibility of eliminating the damping rings in the first stage.  Both
would provide major cost savings.  If the Higgs is discovered at LHC and its
mass is low it might also be worthwhile to create a lower energy dedicated
photon collider Higgs factory as a first stage to the linear collider program.


\begin{footnotesize}


\end{footnotesize}


\end{document}